\documentclass[12pt,a4paper]{iopart}

\usepackage{iopams}
\usepackage{setstack}
\usepackage{graphicx}
\usepackage{hyperref}

\newcommand{\affiliation}[1]{\address{#1}}
\renewcommand{\pacs}[1]{\noindent\textbf{PACS numbers:} #1}
\newcommand{\keywords}[1]{\noindent\textbf{Keywords:} #1}
\renewcommand{\text}[1]{\mathrm{#1}}

\begin{document}

\title[Ground state energy of the $\delta$-Bose and Fermi gas from double extrapolation]{Ground state energy of the $\delta$-Bose and Fermi gas at weak coupling from double extrapolation}
\author{Sylvain Prolhac}
\affiliation{Laboratoire de Physique Th\'eorique, IRSAMC, UPS, Universit\'e de Toulouse, CNRS, France}

\begin{abstract}
We consider the ground state energy of the Lieb-Liniger gas with $\delta$ interaction in the weak coupling regime $\gamma\to0$. For bosons with repulsive interaction, previous studies gave the expansion $e_{\text{B}}(\gamma)\simeq\gamma-4\gamma^{3/2}/3\pi+(1/6-1/\pi^{2})\gamma^{2}$. Using a numerical solution of the Lieb-Liniger integral equation discretized with $M$ points and finite strength $\gamma$ of the interaction, we obtain very accurate numerics for the next orders after extrapolation on $M$ and $\gamma$. The coefficient of $\gamma^{5/2}$ in the expansion is found approximately equal to $-0.00158769986550594498929$, accurate within all digits shown. This value is supported by a numerical solution of the Bethe equations with $N$ particles followed by extrapolation on $N$ and $\gamma$. It was identified as $(3\zeta(3)/8-1/2)/\pi^{3}$ by G. Lang. The next two coefficients are also guessed from numerics. For balanced spin $1/2$ fermions with attractive interaction, the best result so far for the ground state energy was $e_{\text{F}}(\gamma)\simeq\pi^{2}/12-\gamma/2+\gamma^{2}/6$. An analogue double extrapolation scheme leads to the value $-\zeta(3)/\pi^{4}$ for the coefficient of $\gamma^{3}$.\\

\keywords{$\delta$-Bose gas, $\delta$-Fermi gas, ground state energy, Richardson extrapolation}\\
\pacs{02.30.Ik, 02.60.Ed, 02.60.Nm, 05.30.Jp}
\end{abstract}

\maketitle

\begin{section}{Introduction}
The Lieb-Liniger model \cite{LL1963.1} is a one-dimensional quantum integrable model describing $N$ particles with contact interaction. The Hamiltonian of the model is
\begin{equation}
H_{N}=-\sum_{j=1}^{N}\frac{\partial^{2}}{\partial x_{j}^{2}}+2c\sum_{1\leq j<k\leq N}\delta(x_{j}-x_{k})\;.
\end{equation}
Particles are assumed to move on a circle of length $L$, thus $x_{j}\equiv x_{j}+L$. Motivated by recent mathematical results \cite{TW2016.1,TW2016.2,TW2016.3} in the weak coupling regime $|c|\ll1$, we focus on two distinct cases: bosons with repulsive interaction and spin $1/2$ fermions with attractive interaction. For bosons, the Lieb-Liniger model has found recent applications in the context of ultra-cold atomic gases \cite{BDZ2008.1,CCGOR2011.1,YZYYXW2015.1,BF2016.1}. For fermions, the model has been studied in the context of the Bose-Einstein condensation (BEC) of Bardeen-Cooper-Schrieffer (BCS) bound pairs \cite{GPS2008.1,GBC2013.1}.

The Lieb-Liniger model is exactly solvable by Bethe ansatz. For bosons, the ground state energy can be written as $E_{\text{B}}=\sum_{j=1}^{N}q_{j}^{2}$, with $q_{j}$, $j=1,\ldots,N$ solution of the Bethe equations
\begin{equation}
\label{Bethe eqs}
\rmi q_{j}-\frac{1}{L}\sum_{k=1}^{N}\log\Big(-\frac{q_{j}-q_{k}+\rmi c}{q_{j}-q_{k}-\rmi c}\Big)=\frac{2\rmi\pi}{L}\Big(j-\frac{N+1}{2}\Big)\;.
\end{equation}
At large $L,N$ with fixed density of particles $\rho=N/L$, the rescaled eigenvalue $e_{\text{B}}(\gamma)=E_{\text{B}}/(\rho^{2}N)$ with attractive interaction $c>0$ is known to depend on $L$, $N$, $c$ only through the reduced coupling constant $\gamma=c/\rho$. One has then $e_{\text{B}}(\gamma)=\frac{\gamma^{3}}{\kappa_{\text{B}}^{3}}\int_{-1}^{1}\rmd y\,y^{2}f_{\text{B}}(y)$ where $f_{\text{B}}$ is the solution of the Lieb-Liniger integral equation \cite{LL1963.1}
\begin{equation}
\label{Lieb-Liniger integral eq}
f_{\text{B}}(x)=\frac{1}{2\pi}+\frac{\kappa_{\text{B}}}{\pi}\int_{-1}^{1}\rmd y\,\frac{f_{\text{B}}(y)}{(y-x)^{2}+\kappa_{\text{B}}^{2}}\;,
\end{equation}
and $\kappa_{\text{B}}/\gamma=\int_{-1}^{1}\rmd y\,f_{\text{B}}(y)$.

For spin $1/2$ fermions, a nested version of the Bethe ansatz has to be used \cite{Y1967.1} in order to compute the ground state energy $E_{\text{F}}$. At large $L,N$ in the sector with total spin zero, expressions similar to the ones in the bosonic case exist for the rescaled eigenvalue $e_{\text{F}}(\gamma)=\frac{\gamma^{2}}{4}+E_{\text{F}}/(\rho^{2}N)$ with repulsive interaction $c<0$. In terms of the reduced coupling constant $\gamma=|c|/\rho$, one has $e_{\text{F}}(\gamma)=\frac{\gamma^{3}}{\kappa_{\text{F}}^{3}}\int_{-1}^{1}\rmd y\,y^{2}f_{\text{F}}(y)$ where $f_{\text{F}}$ is the solution of the Gaudin integral equation \cite{G1967.1}
\begin{equation}
\label{Gaudin integral eq}
f_{\text{F}}(x)=\frac{2}{\pi}-\frac{\kappa_{\text{F}}}{\pi}\int_{-1}^{1}\rmd y\,\frac{f_{\text{F}}(y)}{(y-x)^{2}+\kappa_{\text{F}}^{2}}\;,
\end{equation}
and $\kappa_{\text{F}}/\gamma=\int_{-1}^{1}\rmd y\,f_{\text{F}}(y)$.

\begin{figure}
  \begin{center}
    \includegraphics[width=150mm]{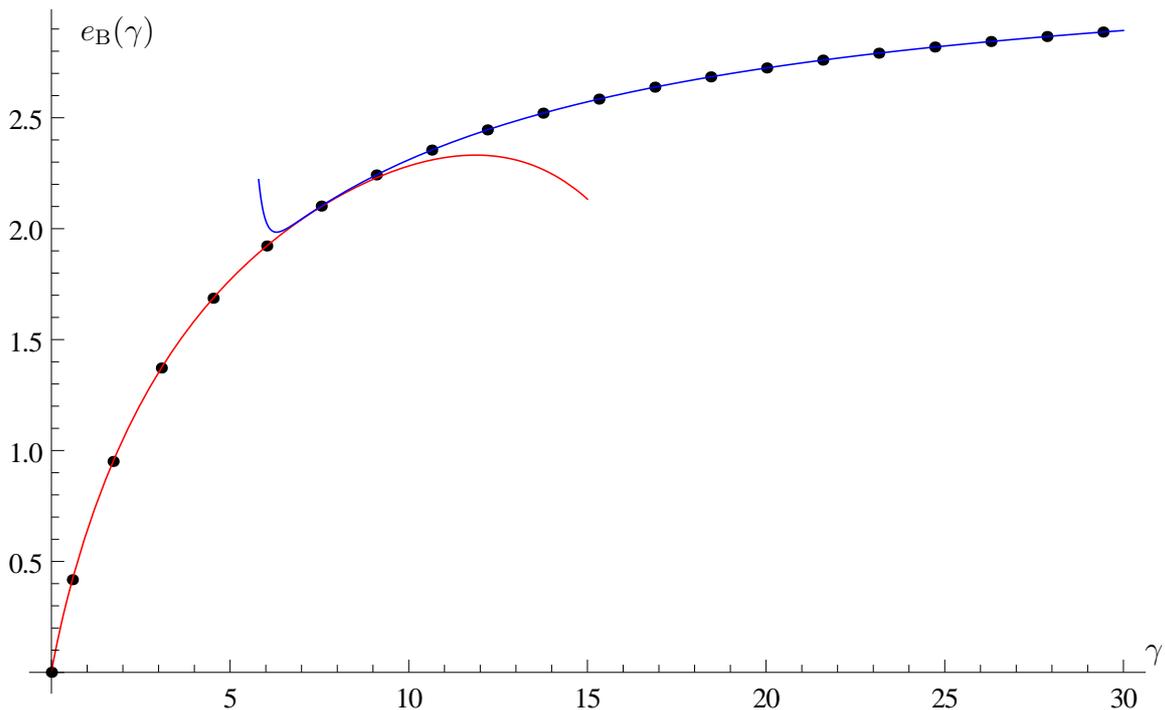}
    \begin{picture}(0,0)
      \put(-2,7){$\gamma$}
      \put(-142,89){$e_{\text{B}}(\gamma)$}
    \end{picture}
  \end{center}
  \caption{Plot of the ground state energy $e_{\text{B}}(\gamma)=E_{\text{B}}/(\rho^{2}N)$ of the $\delta$-Bose gas with repulsive interaction as a function of the strength $\gamma=c/\rho$ of the interaction. The black dots are the result of the extrapolation on the number of particles starting from the integral equation (\ref{Lieb-Liniger integral eq}). The solid red line (left) is the small $\gamma$ expansion $e_{\text{B}}(\gamma)\simeq\sum_{\ell=0}^{9}e_{\ell}^{\text{B}}\gamma^{1+\ell/2}$ with $e_{\ell}^{\text{B}}$'s from table \ref{table extrapolation LL B}. The solid blue line (right) is the large $\gamma$ expansion up to order $\gamma^{-30}$ obtained from the inversion of the integral operator in the Lieb-Liniger equation perturbatively in $1/\gamma$.}
  \label{fig eB}
\end{figure}

\begin{figure}
  \begin{center}
    \includegraphics[width=150mm]{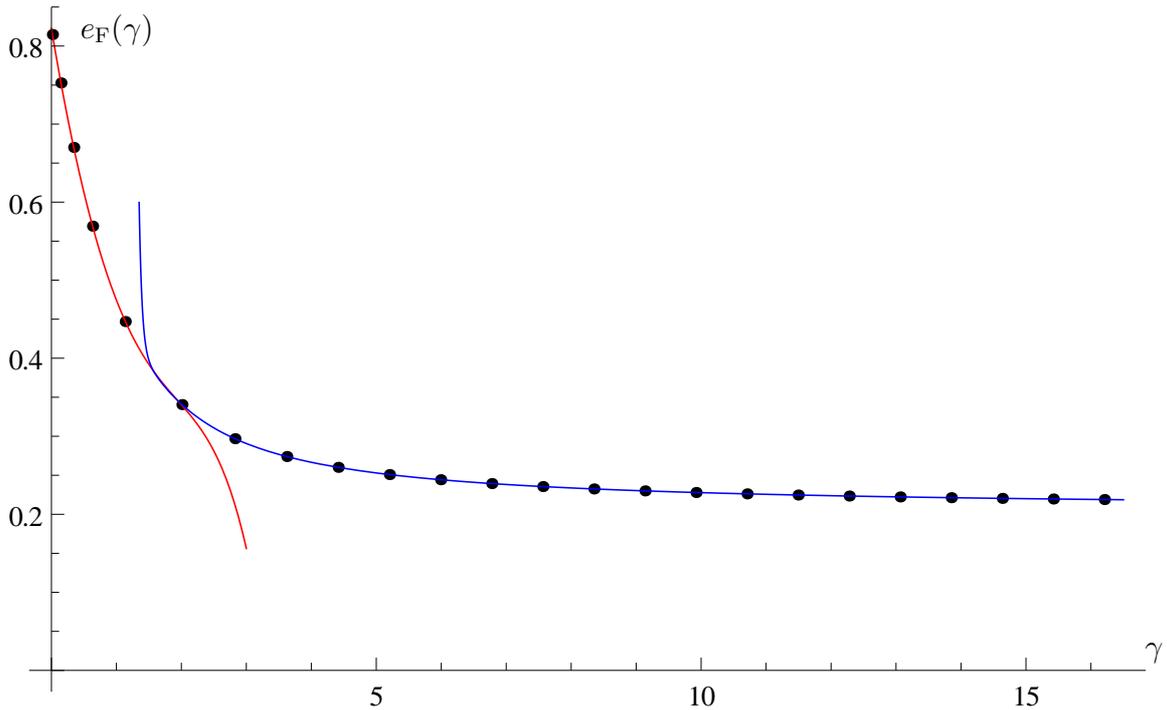}
    \begin{picture}(0,0)
      \put(-2,7){$\gamma$}
      \put(-142,89){$e_{\text{F}}(\gamma)$}
    \end{picture}
  \end{center}
  \caption{Plot of the ground state energy $e_{\text{F}}(\gamma)=\frac{\gamma^{2}}{4}+E_{\text{F}}/(\rho^{2}N)$ of the balanced spin $1/2$ $\delta$-Fermi gas with attractive interaction as a function of the strength $\gamma=|c|/\rho$ of the interaction. The black dots are the result of the extrapolation from the integral equation (\ref{Gaudin integral eq}). The solid red line (left) is the small $\gamma$ expansion $e_{\text{F}}(\gamma)\simeq\sum_{\ell=0}^{6}e_{\ell}^{\text{F}}\gamma^{\ell}$ with $e_{\ell}^{\text{F}}$'s from table \ref{table extrapolation LL F}. The solid blue line (right) is the large $\gamma$ expansion up to order $\gamma^{-30}$ obtained from the inversion of the integral operator in the Gaudin equation perturbatively in $1/\gamma$.}
  \label{fig eF}
\end{figure}

At strong coupling $\gamma\gg1$, the integral equation (\ref{Lieb-Liniger integral eq}) can be solved by inverting the integral operator \cite{TW2016.1}, as the convergent expansion $f_{\text{B}}(x)=\sum_{m=0}^{\infty}(\mathbb{K}^{m}\cdot\frac{1}{2\pi})(x)$ with $(\mathbb{K}\cdot g)(x)=\frac{\kappa_{\text{B}}}{\pi}\int_{-1}^{1}\frac{g(y)}{(y-x)^{2}+\kappa^{2}}$. Since $(\mathbb{K}^{m}\cdot\frac{1}{2\pi})(x)$ is of order $\kappa^{-m}$ when $\kappa\to\infty$, the large $\kappa$ expansion of $f_{\text{B}}(x)$ can be obtained systematically using a computer algebra system by truncating the sum over $m$ and performing explicitly the polynomial integrations resulting from the large $\kappa$ expansion of the kernel. It leads to a large $\gamma$ expansion of $e_{\text{B}}(\gamma)$, where the coefficient of $\gamma^{-m}$, $m\geq1$, is a polynomial in $\pi^{2}$ of order $\lceil m/2\rceil$. The beginning of the expansion is $e_{\text{B}}(\gamma)\simeq\frac{\pi^{2}}{3}-\frac{4\pi^{2}}{3\gamma}+\frac{4\pi^{2}}{\gamma^{2}}+\ldots$, see figure \ref{fig eB} for a plot of the expansion up to order $\gamma^{-30}$, and \cite{R2014.1,LHM2016.1} for a recent study of the structure of the large $\gamma$ expansion. Similarly, in the fermionic case, one finds instead $e_{\text{F}}(\gamma)\simeq\frac{\pi^{2}}{48}+\frac{\pi^{2}}{48\gamma}+\frac{\pi^{2}}{64\gamma^{2}}+\ldots$, see figure \ref{fig eF} for a plot of the expansion up to order $\gamma^{-30}$, and \cite{FRZ2004.1,ZXM2012.1} for studies of the large $\gamma$ expansion in relation with the BCS-BEC crossover.

Weak coupling $\gamma\ll1$ is much harder. In the case of bosons with repulsive interaction, analytical and numerical studies point to an expansion in powers of $\gamma^{1/2}$, known exactly only up to order $\gamma^{2}$ \cite{T1975.1,P1977.1,TW2016.1}:
\begin{equation}
\label{eB[abc]}
e_{\text{B}}(\gamma)\simeq\gamma-\frac{4\gamma^{3/2}}{3\pi}+\Big(\frac{1}{6}-\frac{1}{\pi^{2}}\Big)\gamma^{2}+a\gamma^{5/2}+b\gamma^{3}+c\gamma^{7/2}+\ldots
\end{equation}
Previous numerical evaluations provided the values  $a\approx0.001597$ \cite{L1974.1}, $a\approx0.0018$ \cite{T1975.1}, $a\approx-0.001588$ \cite{EK2001.1}. In the spin $1/2$ Fermi case with attractive interaction, analytical studies indicate an expansion with integer powers of $\gamma$, known exactly only up to order $\gamma^{2}$ \cite{IW2007.1,TW2016.2,TW2016.3}:
\begin{equation}
\label{eF[d]}
e_{\text{F}}(\gamma)\simeq\frac{\pi^{2}}{12}-\frac{\gamma}{2}+\frac{\gamma^{2}}{6}+d\gamma^{3}+\ldots
\end{equation}

Our main results are very precise numerical values for $a$, $b$, $c$ and $d$ obtained using the implementation of Richardson extrapolation \cite{R1927.1} using balanced rational functions, known as the Bulirsch-Stoer method \cite{BS1991.1,HS1988.1}. The method allows to extract the limit $h_{0}=h(0)$ of a function $h(z)$ with series expansion $h(z)\simeq\sum_{k=0}^{\infty}h_{k}z^{\theta k}$ near $z=0$, knowing only a few values $h(z_{n})$, $0<z_{n}<\ldots<z_{1}$ with high numerical precision. The method provides a numerically stable evaluation at $z=0$ of the unique interpolating rational function $R(z)=P(z)/Q(z)$ with $P$ and $Q$ polynomials of respective degrees $\lfloor(n-1)/2\rfloor$ and $\lfloor n/2\rfloor$, such that $R(z_{m})=h(z_{m})$, $m=1,\ldots,n$. The sequence $h(z_{1}),\ldots,h(z_{n})$ is reduced to the extrapolated value $R(0)$ by iteratively building the triangle $h_{m}^{p}$, $p=0,\ldots,n$, $m=1,\ldots,n+1-p$ with the recursion formula
\begin{equation}
\label{Bulirsch-Stoer}
h_{m}^{p}=h_{m+1}^{p-1}+\frac{h_{m+1}^{p-1}-h_{m}^{p-1}}{\big(\frac{z_{m+1}}{z_{m+p+1}}\big)^{\theta}\Big(1-\frac{h_{m+1}^{p-1}-h_{m}^{p-1}}{h_{m+1}^{p-1}-h_{m+1}^{p-2}}\Big)-1}\;,
\end{equation}
and boundary conditions $h_{m}^{0}=0$, $m=1,\ldots,n+1$ and $h_{m}^{1}=h(z_{m})$, $m=1,\ldots,n$. It leads to $h_{1}^{n}=R(0)$. A useful estimator for the order of magnitude of the error is $\varepsilon=|h_{1}^{n}-h_{1}^{n-1}|+|h_{1}^{n}-h_{2}^{n-1}|+|h_{1}^{n-1}-h_{2}^{n-1}|$.

Richardson extrapolation can for instance provide very precise solutions of ordinary differential equations using the modified midpoint approximation scheme \cite{G1965.1} with $2M$ steps, which is known to produce an expansion in $1/M$ with exponent $\theta=2$. We believe the method is also very useful for extracting the thermodynamic limit of observables in quantum integrable models, where one typically finds clean expansions in the inverse of the system size with simple, easy to guess exponents, see \cite{P2016.2} for an application to the asymmetric exclusion process, an exactly solvable lattice model with classical hard-core particles hopping with a preferred direction.

We studied in this paper the ground state energy of the $\delta$-Bose gas at weak coupling using a numerical solution of either the Bethe equations (\ref{Bethe eqs}), see section \ref{section Bethe}, or the integral equation (\ref{Lieb-Liniger integral eq}), see section \ref{section Lieb-Liniger}. Using Richardson extrapolation, we were able to check with very high precision the beginning of the expansion (\ref{eB[abc]}) by subtracting the first terms and dividing by the appropriate power of $\gamma$, see tables \ref{table extrapolation BE B} and \ref{table extrapolation LL B} . Our best result for $a$, coming from the method with the integral equation, is
\begin{equation}
\label{a num}
a\approx-0.00158769986550594498929\;.
\end{equation}
The numerical value has been truncated at the estimator for the error $\varepsilon$. Experience with several other extrapolation calculations, where exact values were also available for comparison, indicates that it is very likely that all the digits displayed in (\ref{a num}) are correct, except the last one that may be off by $1$ due to rounding. The approximate value (\ref{a num}) was then identified as
\begin{equation}
\label{a exact}
a=\Big(\frac{3\zeta(3)}{8}-\frac{1}{2}\Big)\frac{1}{\pi^{3}}
\end{equation}
by G. Lang \cite{L2016.1} shortly after the first version of the paper appeared on the arXiv.

The coefficient $b$ of $e(\gamma)$ can similarly be obtained by the extrapolation method after subtracting (\ref{eB[abc]}) with the exact value (\ref{a exact}). We find
\begin{equation}
\label{b num}
b\approx-0.00016846018782773903545\;,
\end{equation}
roughly in agreement with the result $b\approx-0.000171$ of \cite{EK2001.1}. By analogy with (\ref{a exact}), we sought an exact expression for $\pi^{4}b$ as a linear combination of $1$, $\zeta(2)=\pi^{2}/6$, $\zeta(3)$ and $\zeta(4)=\pi^{4}/90$ with rational coefficients using an integer relations algorithm. We found
\begin{equation}
b=\Big(\frac{\zeta(3)}{8}-\frac{1}{6}\Big)\frac{1}{\pi^{4}}\;,
\end{equation}
in perfect agreement with the numerical value. Iterating again, the next term is numerically equal to
\begin{equation}
\label{c num}
c\approx-0.0000208649733584017408\;,
\end{equation}
for which an exact value is presumably
\begin{equation}
c=\Big(-\frac{45\zeta(5)}{1024}+\frac{15\zeta(3)}{256}-\frac{1}{32}\Big)\frac{1}{\pi^{5}}\;.
\end{equation}
A natural conjecture is that $\pi^{m}$ times the coefficient of $\gamma^{1+m/2}$ is a linear combination of $1$, $\zeta(2)$, \ldots, $\zeta(m)$ with rational coefficients. In that regard, the absence of even zetas for $a$, $b$ and $c$ is slightly odd.

Numerical values of the next coefficients of $e_{\text{B}}(\gamma)$ can be obtained by iterating the procedure further. It seems however difficult to guess other exact numbers as the precision of the numerical values decreases at each step while the requirement for finding exact numbers increases due to the number of zetas involved in the linear combination. We note that using non-exact values in the subtraction quickly degrades the quality of both the extrapolated result and the estimation for the error (which is typically underestimated) as numerical errors accumulate.

We also applied the extrapolation procedure in the Fermi case, starting with the integral equation (\ref{Gaudin integral eq}). Again, we were able to check (\ref{eF[d]}) with high precision, see table \ref{table extrapolation LL F}. We found for the coefficient $d$ the numerical value
\begin{equation}
\label{d num}
d\approx-0.0123402948369572\;,
\end{equation}
which agrees perfectly with the exact number
\begin{equation}
d=-\frac{\zeta(3)}{\pi^{4}}\;.
\end{equation}
Numerics for the next coefficients are displayed in table \ref{table extrapolation LL F}. We were not able to determine other exact values for them, due to the relatively small number of digits known. A smaller number of coefficients is accessible with the same amount of computational effort in the Fermi case compared to the Bose case. This is a consequence of the fact that the expansion is in integer powers of $\gamma$ for the Fermi gas instead of $\gamma^{1/2}$ for the Bose gas, and thus requires more accurate values before extrapolating on $\gamma$.

A motivation for the present work were similar numerical extrapolations performed in \cite{P2016.2} for the spectral gap of a weakly asymmetric exclusion process, an exactly solvable lattice model featuring classical hard-core particles hopping with a slightly preferred direction between neighbouring sites of a one-dimensional lattice. In that case too, it had been possible to guess exact expressions for the coefficients from extrapolated numerical values.
\end{section}

\begin{section}{Double extrapolation from Bethe equations at finite number of particles}
\label{section Bethe}
In this section, we provide independent support of the numerical value (\ref{a num}), (\ref{b num}) and (\ref{c num}) obtained with higher precisely in section \ref{section Lieb-Liniger}, by solving numerically the Bethe equations (\ref{Bethe eqs}) and extrapolating on the number of particles.

\begin{table}
  \begin{center}
    \begin{tabular}{rlc}
     $m$ & \hspace{1mm}$\hspace{3mm}\displaystyle e_{m}^{\text{B}}=\lim_{\gamma\to0}\frac{1}{\gamma^{1+m/2}}\Big(e_{\text{B}}(\gamma)-\sum_{\ell=0}^{m-1}e_{\ell}^{\text{B}}\,\gamma^{1+\ell/2}\Big)$ & \begin{tabular}{c}relative error from\vspace{-1mm}\\the exact value\end{tabular}\vspace{2mm}\\
     0 & $\hspace{3.2mm}1.0000000000000000000\hspace{5mm}\approx1$ & $<10^{-20}$\\
     1 & $-0.42441318157838756205\hspace{2.8mm}\approx-4/3\pi$ & $<10^{-20}$\\
     2 & $\hspace{3.2mm}0.0653454830243288952\hspace{5mm}\approx1/6-1/\pi^{2}$ & $<10^{-19}$\\
     3 & $-0.00158769986550594499\hspace{2.8mm}\approx a$ & $<10^{-18}$\\
     4 & $-0.0001684601878277390\hspace{4.9mm}\approx b$ & $<10^{-16}$\\
     5 & $-0.000020864973358402\hspace{7mm}\approx c$ & $<10^{-15}$\\
     6 & $-3.1632142185374\times10^{-6}$ &\\
     7 & $-6.10686059567\times10^{-7}$ &\\
     8 & $-\underline{1.4840346726}50\times10^{-7}$ &\\
     9 & $-\underline{4.3460962}4646\times10^{-8}$ &\\
     10 & $-\underline{1.474253}312\times10^{-8}$ &
    \end{tabular}
  \end{center}
  \caption{Result of the extrapolation for the ground state energy $e_{\text{B}}(\gamma)\simeq\sum_{m=0}^{\infty}e_{m}^{\text{B}}\,\gamma^{1+m/2}$ of the $\delta$-Bose gas at weak coupling from a numerical solution of the Bethe equations for a finite number of particles. The values are truncated at the estimation of the error $\varepsilon$, see below (\ref{Bulirsch-Stoer}). For $m=0,\ldots,7$ they are accurate to the number of digits displayed. For $m\geq8$ they are accurate only to the number of digits underlined.}
  \label{table extrapolation BE B}
\end{table}

Using Newton's method, we obtain numerical values of the Bethe roots $q_{j}$, $j=1,\ldots,N$ satisfying the Bethe equations for the ground state (\ref{Bethe eqs}) by solving iteratively the linear system
\begin{eqnarray}
&& \sum_{k=1}^{N}(q_{k}^{\text{new}}-q_{k})\Big[\rmi\delta_{j,k}+\frac{1}{L}\Big(\frac{1}{q_{j}-q_{k}+\rmi c}-\frac{1}{q_{j}-q_{k}-\rmi c}\Big)\nonumber\\
&&\hspace{33mm} -\frac{\delta_{j,k}}{L}\sum_{\ell=1}^{N}\Big(\frac{1}{q_{j}-q_{\ell}+\rmi c}-\frac{1}{q_{j}-q_{\ell}-\rmi c}\Big)\Big]\\
&& =-\rmi q_{j}+\frac{2\rmi\pi}{L}\Big(j-\frac{N+1}{2}\Big)+\frac{1}{L}\sum_{k=1}^{N}\log\Big(-\frac{q_{j}-q_{k}+\rmi c}{q_{j}-q_{k}-\rmi c}\Big)\;\nonumber
\end{eqnarray}
for $q_{k}^{\text{new}}$, $k=1,\ldots,N$. We start with the known solution for $Lc\to0$ with fixed $N$, where $q_{j}$, $j=1,\ldots,N$ are given in terms of the roots of the $N$-th Hermite polynomial: $H_{N}(\sqrt{2cq_{j}/L})=0$ with $H_{N}(x)=(-1)^{N}\rme^{x^{2}}\partial_{x}^{N}\rme^{-x^{2}}$. The calculations were done at density $\rho=1$, with $1000$ significant digits, for all integers $N$ between $1$ and $N_{\text{max}}(\gamma)$, and for the $100$ values $\gamma=1,0.99,\ldots,0.02,0.01$. The integers $N_{\text{max}}(\gamma)$ chosen increase from $77$ to $308$ as $\gamma$ decreases from $1$ to $0.01$ in order to maintain a comparable precision for all $\gamma$ after the extrapolation on the number of particles.

For each value of $\gamma$, the limit $N\to\infty$ is reached by extrapolating on $1/N$ with exponent $\theta=2$, chosen among simple values to minimize the estimation for the error. The limit $\gamma\to0$ is then obtained by extrapolating on $\gamma$, with exponent $\theta=1/2$ imposed by the form of the weak coupling expansion. The results are summarized in table \ref{table extrapolation BE B}. Perfect agreement is found with known analytical results.
\end{section}

\begin{section}{Double extrapolation from the integral equations}
\label{section Lieb-Liniger}
In this section, we explain the computations leading to (\ref{a num}), (\ref{b num}) and (\ref{c num}) starting with a solution of a discretized version of the Lieb-Liniger integral equation (\ref{Lieb-Liniger integral eq}) and extrapolating on the number of points of discretization. We also treat the fermionic case from the Gaudin integral equation (\ref{Gaudin integral eq}).

\begin{table}
  \begin{center}
    \begin{tabular}{rlc}
     $m$ & \hspace{5mm}$\displaystyle e_{m}^{\text{B}}=\lim_{\gamma\to0}\frac{1}{\gamma^{1+m/2}}\Big(e_{\text{B}}(\gamma)-\sum_{\ell=0}^{m-1}e_{\ell}^{\text{B}}\,\gamma^{1+\ell/2}\Big)$ & \begin{tabular}{c}relative error from\vspace{-1mm}\\the exact value\end{tabular}\vspace{2mm}\\
     0 & $\hspace{3.2mm}1.0000000000000000000000001\approx1$ & $<10^{-24}$\\
     1 & $-0.4244131815783875620504\hspace{6mm}\approx-4/3\pi$ & $<10^{-22}$\\
     2 & $\hspace{3.2mm}0.06534548302432889522279\hspace{4mm}\approx1/6-1/\pi^{2}$ & $<10^{-23}$\\
     3 & $-0.00158769986550594498929\hspace{4mm}\approx a$ & $<10^{-21}$\\
     4 & $-0.00016846018782773903545\hspace{4mm}\approx b$ & $<10^{-20}$\\
     5 & $-0.0000208649733584017408\hspace{6mm}\approx c$ & $<10^{-18}$\\
     6 & $-3.16321421853736662\times10^{-6}$ &\\
     7 & $-\underline{6.1068605956750}2\times10^{-7}$ &\\
     8 & $-\underline{1.48403467261}867\times10^{-7}$ &\\
     9 & $-\underline{4.3460962537}10042\times10^{-8}$ &\\
     10 & $-\underline{1.474253}1944024322\times10^{-8}$ &
    \end{tabular}
  \end{center}
  \caption{Result of the double extrapolation for the ground state energy $e_{\text{B}}(\gamma)\simeq\sum_{m=0}^{\infty}e_{m}^{\text{B}}\,\gamma^{1+m/2}$ of the repulsive $\delta$-Bose gas at weak coupling from a numerical solution of the Lieb-Liniger integral equation (\ref{Lieb-Liniger integral eq}). The values are truncated at the estimation of the error $\varepsilon$, see below (\ref{Bulirsch-Stoer}). For $m=0,\ldots,6$ they are accurate to the number of digits displayed. For $m\geq7$ they are roughly accurate to the number of digits underlined.}
  \label{table extrapolation LL B}
\end{table}

In the bosonic case, we discretize the Lieb-Liniger integral equation (\ref{Lieb-Liniger integral eq}) for several values of $\kappa$ by replacing the function $f_{\text{B}}$ by the list of $M$ values $f(x_{m})$ with $x_{m}=-1+2(m-1)/(M-1)$, $m=1,\ldots,M$. The integral in (\ref{Lieb-Liniger integral eq}) is replaced by a trapezoidal approximation. The solution of the discretized integral equation is then computed with $1000$ digits by solving for $f(x_{m})$, $m=1,\ldots,M$ the linear system
\begin{equation}
\fl\hspace{0mm} f(x_{m})=\frac{1}{2\pi}+\frac{\kappa}{\pi}\Big(\frac{1}{2}\,\frac{f(x_{1})}{(x_{1}-x_{m})^{2}+\kappa^{2}}+\sum_{n=2}^{M-1}\frac{f(x_{n})}{(x_{n}-x_{m})^{2}+\kappa^{2}}+\frac{1}{2}\,\frac{f(x_{M})}{(x_{M}-x_{m})^{2}+\kappa^{2}}\Big)\;.
\end{equation}
The reduced coupling $\gamma_{M}$ is then computed from
\begin{equation}
\gamma_{M}=\kappa\Big/\Big(\frac{f(x_{1})}{2}+\sum_{n=2}^{M-1}f(x_{n})+\frac{f(x_{M})}{2}\Big)\;,
\end{equation}
and the approximate value of the ground state energy is given by
\begin{equation}
e_{M}=\frac{\gamma_{M}^{3}}{\kappa^{3}}\Big(\frac{x_{1}^{2}\,f(x_{1})}{2}+\sum_{n=2}^{M-1}x_{n}^{2}\,f(x_{n})+\frac{x_{M}^{2}\,f(x_{M})}{2}\Big)\;.
\end{equation}

This process is repeated with $M=2,3,\ldots,M_{\text{max}}(\kappa)$ for each value of $\kappa$. Extrapolating on $M$ with exponent $\theta=1$ leads to accurate values for $e_{\text{B}}(\gamma)$ with $\gamma$ given implicitly in terms of $\kappa$. This is done for the $100$ values $\kappa=1,0.99,\ldots,0.02,0.01$, with $M_{\text{max}}(\kappa)$ increasing from $71$ to $803$ as $\kappa\to0$ (which corresponds to $\gamma\to0$). Extrapolation on $\kappa$ with exponent $\theta=1/2$ then leads to the results in table \ref{table extrapolation LL B}. Again, perfect agreement is found with known analytical results. The weak coupling expansion is plotted from table \ref{table extrapolation LL B} in figure \ref{fig eB}.

\begin{table}
  \begin{center}
    \begin{tabular}{rlc}
     $m$ & $\displaystyle e_{m}^{\text{F}}=\lim_{\gamma\to0}\frac{1}{\gamma^{m}}\Big(e_{\text{F}}(\gamma)-\sum_{\ell=0}^{m-1}e_{\ell}^{\text{F}}\,\gamma^{\ell}\Big)$ & \begin{tabular}{c}relative error from\vspace{-1mm}\\the exact value\end{tabular}\vspace{2mm}\\
     0 & $\hspace{3.2mm}0.82246703342411322\approx\pi^{2}/12$ & $<10^{-18}$\\
     1 & $-0.50000000000000000\approx-1/2$ & $<10^{-18}$\\
     2 & $\hspace{3.2mm}0.1666666666666667\hspace{2mm}\approx1/6$ & $<10^{-17}$\\
     3 & $-0.0123402948369572\hspace{2mm}\approx-\zeta(3)/\pi^{4}$ & $<10^{-15}$\\
     4 & $-0.001875499919064$ &\\
     5 & $-\underline{0.00038005574}3439$ &\\
     6 & $-\underline{0.000121746}636000$ &
    \end{tabular}
  \end{center}
  \caption{Result of the double extrapolation for the ground state energy $e_{\text{F}}(\gamma)\simeq\sum_{m=0}^{\infty}e_{m}^{\text{F}}\,\gamma^{m}$ of the attractive $\delta$-Fermi gas with spin $1/2$ at weak coupling from a numerical solution of the Lieb-Liniger integral equation (\ref{Gaudin integral eq}). The values are truncated at the estimation of the error $\varepsilon$, see below (\ref{Bulirsch-Stoer}). For $m=0,\ldots,4$ they are accurate to the number of digits displayed. For $m\geq5$ they are roughly accurate to the number of digits underlined.}
  \label{table extrapolation LL F}
\end{table}

A similar double extrapolation scheme is used for the Fermi gas starting from the integral equation (\ref{Gaudin integral eq}). The extrapolation on $\kappa$ is done for the $100$ values $\kappa=1,0.99,\ldots,0.02,0.01$, with $M_{\text{max}}(\kappa)$ increasing from $67$ to $771$ as $\kappa\to0$. The biggest difference is that the exponent $\theta=1$ is used for the extrapolation on $\kappa$ due to the form of the expansion (\ref{eF[d]}). The results, given in table \ref{table extrapolation LL F}, are in complete agreement with (\ref{eF[d]}). The weak coupling expansion is plotted from table \ref{table extrapolation LL F} in figure \ref{fig eF}.
\end{section}

\begin{section}{Conclusions}
We computed in this paper accurate values for the coefficients of the ground state energies $e_{\text{B}}(\gamma)$ and $e_{\text{F}}(\gamma)$ of the $\delta$-Bose gas and spin $1/2$ $\delta$-Fermi gas in the weak coupling expansion using Richardson extrapolation. The efficiency of the extrapolation method for this problem, measured by the very small value for the estimator of the error, is a sign that $e_{\text{B}}(\gamma)$ and $e_{\text{F}}(\gamma)$ have clean expansions respectively in powers of $\gamma^{1/2}$ and $\gamma$. In particular, it is a good indication for the absence of sub-leading logarithm in the expansions, since logarithms are known to completely spoil the precision of Richardson extrapolation. This is in agreement with exact calculations \cite{TW2016.1,TW2016.3} showing the cancellation of logarithms appearing in intermediate quantities when computing the first few terms of the weak coupling expansion of the ground state energy, but a proof at all orders is still missing.

The very precise numerical results of this paper are another example of the spectacular effectiveness of Richardson extrapolation. For integrable models, it can provide very accurate numerics in the thermodynamic limit knowing only relatively few finite size values, thanks to the existence of clean expansions with simple exponents combined with efficient ways to compute accurate finite size values for a moderately large number of degrees of freedom. Such precise numerics even allow in some cases to guess corresponding exact numbers.\\

\noindent
\textbf{Acknowledgements}: I thank G. Lang for allowing me to use his conjecture for the exact value of $a$ and C. Tracy for suggesting I apply Richardson extrapolation to the Fermi case too. I also thank the hospitality of H. Spohn at TU M\"unchen, where part of this paper was written.
\end{section}

\vspace{10mm}


\begin{thebibliography}{10}

\bibitem{LL1963.1}
E.H. Lieb and W.~Liniger.
\newblock Exact analysis of an interacting {B}ose gas. {I}. the general
  solution and the ground state.
\newblock {\em Phys. Rev.}, 130:1605--1616, 1963.

\bibitem{TW2016.1}
C.A. Tracy and H.~Widom.
\newblock On the ground state energy of the $\delta$-function {B}ose gas.
\newblock {\em J. Phys. A: Math. Theor.}, 49:294001, 2016.

\bibitem{TW2016.2}
C.A. Tracy and H.~Widom.
\newblock On the ground state energy of the delta-function {F}ermi gas.
\newblock {\em J. Math. Phys.}, 57:103301, 2016.

\bibitem{TW2016.3}
C.A. Tracy and H.~Widom.
\newblock On the ground state energy of the delta-function {F}ermi gas {II}:
  Further asymptotics.
\newblock {\em arXiv:1609.07793}, 2016.

\bibitem{BDZ2008.1}
I.~Bloch, J.~Dalibard, and W.~Zwerger.
\newblock Many-body physics with ultracold gases.
\newblock {\em Rev. Mod. Phys.}, 80:885, 2008.

\bibitem{CCGOR2011.1}
M.A. Cazalilla, R.~Citro, T.~Giamarchi, E.~Orignac, and M.~Rigol.
\newblock One dimensional bosons: From condensed matter systems to ultracold
  gases.
\newblock {\em Rev. Mod. Phys.}, 83:1405, 2011.

\bibitem{YZYYXW2015.1}
J.~Yu-Zhu, C.~Yang-Yang, and G.~Xi-Wen.
\newblock Understanding many-body physics in one dimension from the
  {L}ieb-{L}iniger model.
\newblock {\em Chinese Physics B}, 24:050311, 2015.

\bibitem{BF2016.1}
M.T. Batchelor and A.~Foerster.
\newblock {Y}ang-{B}axter integrable models in experiments: from condensed
  matter to ultracold atoms.
\newblock {\em J. Phys. A: Math. Theor.}, 49:173001, 2016.

\bibitem{GPS2008.1}
S.~Giorgini, L.P. Pitaevskii, and S.~Stringari.
\newblock Theory of ultracold atomic {F}ermi gases.
\newblock {\em Rev. Mod. Phys.}, 80:1215, 2008.

\bibitem{GBC2013.1}
X.-W. Guan, M.T. Batchelor, and C.~Lee.
\newblock Fermi gases in one dimension: From {B}ethe ansatz to experiments.
\newblock {\em Rev. Mod. Phys.}, 85:1633, 2013.

\bibitem{Y1967.1}
C.N. Yang.
\newblock Some exact results for the many-body problem in one dimension with
  repulsive delta-function interaction.
\newblock {\em Phys. Rev. Lett.}, 19:1312, 1967.

\bibitem{G1967.1}
M.~Gaudin.
\newblock Un syst\`eme \`a une dimension de fermions en interaction.
\newblock {\em Phys. Lett. A}, 24:55--56, 1967.

\bibitem{R2014.1}
Z.~Ristivojevic.
\newblock Excitation spectrum of the {L}ieb-{L}iniger model.
\newblock {\em Phys. Rev. Lett.}, 113:015301, 2014.

\bibitem{LHM2016.1}
G.~Lang, F.~Hekking, and A.~Minguzzi.
\newblock Ground-state energy and excitation spectrum of the {L}ieb-{L}iniger
  model: accurate analytical results and conjectures about the exact solution.
\newblock {\em arXiv:1609.08865}, 2016.

\bibitem{FRZ2004.1}
J.N. Fuchs, A.~Recati, and W.~Zwerger.
\newblock Exactly solvable model of the {BCS}-{BEC} crossover.
\newblock {\em Phys. Rev. Lett.}, 93:090408, 2004.

\bibitem{ZXM2012.1}
L.~Zhou, C.-Y. Xu, and Y.-L. Ma.
\newblock Exact studies of ground and excited states of one-dimensional
  $\delta$-interacting {F}ermi gases in the {BCS}-{BEC} crossover.
\newblock {\em J. Stat. Mech.}, 2012:L03002, 2012.

\bibitem{T1975.1}
M.~Takahashi.
\newblock On the validity of collective variable description of {B}ose systems.
\newblock {\em Prog. Theor. Phys.}, 53:386--399, 1975.

\bibitem{P1977.1}
V.N. Popov.
\newblock Theory of one-dimensional {B}ose gas with point interaction.
\newblock {\em Theor. Math. Phys.}, 30:222--226, 1977.

\bibitem{L1974.1}
D.K. Lee.
\newblock Ground state of a one-dimensional many-boson system.
\newblock {\em Phys. Rev. A}, 9:1760, 1974.

\bibitem{EK2001.1}
T.~Emig and M.~Kardar.
\newblock Probability distributions of line lattices in random media from the
  {1D} {B}ose gas.
\newblock {\em Nucl. Phys. B}, 604:479--510, 2001.

\bibitem{IW2007.1}
T.~Iida and M.~Wadati.
\newblock Exact analysis of a $\delta$-function spin-1/2 attractive {F}ermi gas
  with arbitrary polarization.
\newblock {\em J. Stat. Mech.}, 2007:P06011, 2007.

\bibitem{R1927.1}
L.F. Richardson.
\newblock The deferred approach to the limit.
\newblock {\em Phil. Trans. R. Soc. A}, 226:636--646, 1927.

\bibitem{BS1991.1}
R.~Bulirsch and J.~Stoer.
\newblock {\em Introduction to Numerical Analysis}.
\newblock New York: Springer-Verlag, 1991.

\bibitem{HS1988.1}
M.~Henkel and G.M. Sch\"utz.
\newblock Finite-lattice extrapolation algorithms.
\newblock {\em J. Phys. A: Math. Gen.}, 21:2617--2633, 1988.

\bibitem{G1965.1}
W.B. Gragg.
\newblock On extrapolation algorithms for ordinary initial value problems.
\newblock {\em Journal of the Society for Industrial and Applied Mathematics:
  Series B, Numerical Analysis}, 2:384--403, 1965.

\bibitem{P2016.2}
S.~Prolhac.
\newblock Extrapolation methods and {B}ethe ansatz for the asymmetric exclusion
  process.
\newblock {\em J. Phys. A: Math. Theor.}, 49:454002, 2016.

\bibitem{L2016.1}
G. Lang.
\newblock {Private communication}, 2016.

\end{thebibliography}

\end{document}